\documentclass{PoS}
\usepackage{bm}

\title{Role and Properties of Wilson Lines in Transverse-Momentum-Dependent
       Parton Distribution Functions}

\ShortTitle{Wilson lines in TMD PDFs}

\author{\speaker{N.~G.~Stefanis}\thanks{Also at Bogoliubov Laboratory of
        Theoretical Physics, JINR, 141980 Dubna, Russia.}\\
        Institut f\"{u}r Theoretische Physik II,
        Ruhr-Universit\"{a}t Bochum,
        D-44780 Bochum, Germany\\
        E-mail: \email{stefanis@tp2.ruhr-uni-bochum.de}}

\author{I.~O.~Cherednikov\thanks{Also at ITPM, Moscow State University,
        119899 Moscow, Russia.}\\
        INFN Cosenza, Universit$\grave{a}$
        della Calabria, I-87036 Rende (CS), Italy and \\
        Bogoliubov Laboratory of Theoretical Physics,
        JINR, 141980 Dubna, Russia\\
        E-mail: \email{igor.cherednikov@jinr.ru}}
\author{A.~I.~Karanikas\\
        University of Athens, Department of Physics,
        Nuclear and Particle Physics Section,
        Panepistimiopolis, GR-15771 Athens, Greece\\
        E-mail: \email{akaran@phys.uoa.gr}}

\abstract{We summarize the renormalization-group properties of
          transverse-momentum dependent (TMD) parton distribution
          functions (PDF)s arguing that in the light-cone gauge
          the overlapping ultraviolet and rapidity divergences
          cannot be solely controlled by (dimensional)
          regularization, but necessitate their renormalization.
          In doing so, we show that at the one-loop order this
          additional divergence entails an anomalous dimension which
          can be attributed to a cusp in the gauge contour at
          light-cone infinity.
          Then, we present a recent analysis of TMD PDFs which
          incorporates in the gauge links the Pauli term
          $\sim F^{\mu\nu}[\gamma_\mu, \gamma_\nu]$.
          This generalized treatment of gauge invariance is shown to
          be justified, in the sense that it does not modify the
          behavior of the leading-twist contribution, though it
          contributes to the anomalous dimension of that of
          twist-three.
          An important consequence of the inclusion of the
          spin-dependent Pauli term is the appearance of a constant
          phase---the same for the leading twist-two and subleading
          distribution functions---that ensues from the interaction
          of the Pauli term in the transverse gauge link with the
          gauge field accompanying the fermion.
          Remarkably, this phase has opposite sign for the Drell-Yan
          process as compared to the semi-inclusive DIS.
          }

\FullConference{Light Cone 2010 - LC2010\\
		June 14-18, 2010\\
		Valencia, Spain}

\begin{document}

\section{Introduction and Theoretical Framework}
\label{sec:intro}

One of the problems inherent in the definition of hadronic observables
is how to ensure gauge invariance.
This problem arises because correlators are nonlocal quantities which
contain local operators that transform differently under gauge
transformations, hence entailing a dependence on the gauge adopted.
Clearly, physical quantities should not depend on the choice of the
gauge we choose to work --- this should merely be a matter of
(calculational) convenience.
To render \emph{integrated} parton distribution functions (PDF)s gauge
invariant, it is sufficient to insert into their definition a Wilson
line --- a gauge link --- between the two Heisenberg quark operators
that renders their product gauge invariant \cite{CS81}.
From the point of view of renormalizability, this operation introduces
additional contributions to the anomalous dimension of the PDFs.
These contributions stem from the local obstructions of the gauge
contours: endpoints, cusps, and self-crossing points
(see \cite{CS08} for a technical exposition and references).
It is to be emphasized that although the gauge link is nonlocal, no
explicit path dependence is introduced, e.g., on the gauge-contour
length.
Actually, to ensure the gauge invariance of the PDFs it is even
sufficient to use a straight lightlike line, because integrated PDFs
are defined on the light cone and the only contribution from the
gauge link to the anomalous dimension of the PDF comes from its
endpoints (see, for instance, \cite{Ste83} and references cited
therein).
Hence, one has for the integrated PDF of a quark $i$ in a quark $a$
\begin{eqnarray}
  f_{i/a}(x)
=
  \frac{1}{2} \int \frac{d\xi^{-}}{2\pi}
  e^{- i k^{+} \xi^{-}}
  \langle P |\bar{\psi}_i (\xi^{-},
  \mathbf{0}_\perp) \gamma^{+}
  [\xi^-, 0^-|\mathcal{C}]
  \psi_{i}(0^{-},\mathbf{0}_\perp) | P
  \rangle \ ,
\label{eq:int-pdf}
\end{eqnarray}
where
\begin{eqnarray}
  [\xi^-, 0^-|\mathcal{C}]
=
 {\cal P} \exp \Bigg[-i g \int_{0^{-}[\mathcal{C}]}^{\xi^-} dz^\mu
        A_{\mu}^{a}(0,z^-, \mathbf{0}_\perp) t_{a}
  \Bigg]
\label{eq:link}
\end{eqnarray}
is a path-ordered gauge link (Wilson line) in the lightlike direction
from $0$ to $\xi$ along the contour~$\mathcal{C}$.

One may insert a complete set of states and split the gauge link
$[\xi^-, 0^-]$ into two gauge links connecting the points $0$ and $\xi$
through $\infty$.
This is mathematically sound, provided the junction (hidden at
infinity) of the two involved contours is smooth, i.e., entails only a
trivial renormalization of the junction point so that the validity of
the algebraic identity
$
  [x_2,z \ |\ \mathcal{C}_1] \ [z,x_1\ | \ \mathcal{C}_2]
=
  [x_2,x_1\ | \ \mathcal{C}=\mathcal{C}_1\cup \mathcal{C}_2]
$
is ensured.
This being the case, it is possible to associate each of the quark
fields with its own gauge link because the attached contour has no
bearing on the definition of $f_{i/a}(x)$.
Then, the struck quark can be replaced by an ``eikonalized quark''
\begin{equation}
  \Psi (x^-|\Gamma)
=
  \psi(x^-) [x^-,\infty^-|\Gamma]
\equiv
  \psi(x^-) \mathcal{P}{\rm exp}
  \left[ -ig \int_{\infty^- [\Gamma]}^{x^-} dz_{\mu}
  A^{\mu}_{a}(0^+, z^-, \mathbf{0}_\perp) t_{a} \right]
\label{eq:eikonal-quark}
\end{equation}
which is a contour-dependent Mandelstam fermion field \cite{Man68YM}
(with an analogous definition for the antifermion field).
In this scheme, the gluon reconstitution in the gauge-invariant
correlator for the integrated PDF involves gluons emanating either
from the gauge links --- giving rise to selfenergy-like diagrams
--- or contractions with the gluon self-fields of the Heisenberg
operator for the struck quark which generate crosstalk-type diagrams.
Note that for the sake of clarity and simplicity, we ignore bound
states (spectators).
As a result, one has
\begin{eqnarray}
  f_{i/a}^{\rm split}(x)
=
  \frac{1}{2} \sum_n \int \frac{d\xi^-}{2\pi}
  e^{- i k^{+} \xi^{-}}
  \langle P |\bar{\Psi}_{i}\left(\xi^-,\mathbf{0}_\perp|{\cal C}_{1}\right)
       | n \rangle \gamma^{+}
  \langle n |\Psi_{i}\left(0^-,\mathbf{0}_\perp|{\cal C}_{2}\right)
       |P\rangle \ .
\label{eq:pdf-link-split}
\end{eqnarray}

This concept was carried over to \emph{unintegrated} PDFs, i.e., to
those PDFs which still depend on the transverse
momenta --- hence termed TMD PDFs.
However, TMD PDFs with only longitudinal gauge links are not completely
gauge invariant under different boundary conditions on the gluon
propagator in the light-cone gauge $A^+=0$.
The reason is that $x^-$-independent gauge transformations are still
possible under the same gauge condition.
Hence, the naive collinear gauge-invariant TMD PDF definition as for
the integrated case is inapplicable.
Refurbishment is provided via the introduction of transverse gauge
links which necessarily stretch out off the light cone to infinity
\cite{BJY03,BMP03}.
This generalizes Eq.\ (\ref{eq:pdf-link-split}) for a quark with
$k_\mu = (k^+, k^-, \mathbf{k}_\perp)$
in a quark with $p_\mu = (p^+, p^-, \mathbf{0}_\perp)$ to the expression
\begin{eqnarray}
  f_{q/q}(x, \mathbf{k}_\perp)
=
   \frac{1}{2}
   \int \frac{d\xi^{-}}{2\pi} \frac{d^2{\bm \xi}_\perp}{(2\pi)^2}
   \exp\left(
        - i k^{+} \xi^{-} + i \mathbf{k}_\perp \cdot {\bm \xi}_\perp
        \right)
   \Big\langle
   q(p) |\bar \psi (\xi^-, {\bm \xi}_\perp)
   [\xi^-, {\bm \xi}_\perp;
   \infty^-, {\bm \xi}_\perp]^{\dagger}\nonumber \\
\times
   [\infty^-, {\bm \xi}_\perp;
   \infty^-, {\bm \infty}_\perp]^\dagger
   \gamma^+
   [\infty^-, {\bm \infty}_\perp;
   \infty^-, \mathbf{0}_\perp]
   [\infty^-, \mathbf{0}_\perp;0^-,
   \mathbf{0}_\perp]
   \psi (0^-, \mathbf{0}_\perp) |q(p)
   \Big\rangle
   \Big|_{\xi^+ =0}
\label{eq:tmd-pdf}
\end{eqnarray}
in which
\begin{equation}
  [\infty^{-}, {\bm \xi}_{\perp}; \xi^{-}, {\bm \xi}_{\perp}]
\equiv
  \mathcal{P} \exp \left[
                         i g \int_{0}^{\infty} d\tau \, n_{\mu}^{-}
                         A_{a}^{\mu} t^{a} (\xi + n^- \tau)
                   \right] \ ,
\label{eq:lightlike-link}
\end{equation}
\begin{equation}
  [
   \infty^{-}, {\bm \infty}_{\perp}; \infty^{-}, {\bm \xi}_{\perp}
  ]
\equiv
 \mathcal{P} \exp \left[
                        i g \int_{0}^{\infty} d\tau \, \mathbf{l}
                        \cdot \mathbf{A}_{a} t^{a}
                        ({\bm \xi}_{\perp} + \mathbf{l}\tau)
                  \right]
\label{eq:transverse-link}
\end{equation}
are the lightlike and the transverse gauge link, respectively.

\section{One-Loop Gluon Virtual Corrections in the $A^+=0$ Gauge}
\label{sec:one-loop-corr}

\begin{figure}
\includegraphics[width=.5\textwidth]{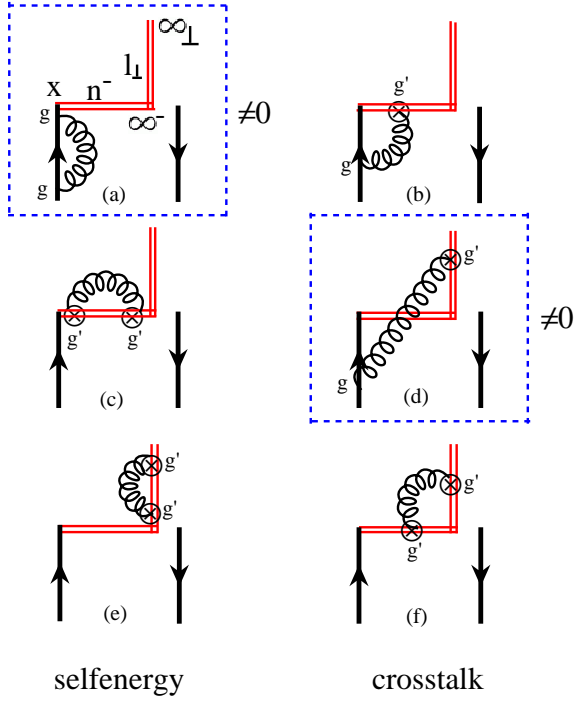}
\caption{One-loop gluon virtual corrections to $f_{q/q}$ in the
$A^+=0$ gauge.
The double lines describe the gauge links attached to the fermions
(heavy lines), while the curly lines represent gluons, and
the symbol $\otimes$ denotes a line integral.
The Hermitian-conjugate (mirror) diagrams are not shown.}
\label{fig:fig1}
\end{figure}

The pursuit of a proper definition of TMD PDFs is a long-standing
problem that was not accomplished with the definition above.
The reason is --- frankly speaking --- that nobody knows how the
contour behaves at light-cone infinity when it ventures out in
the transverse directions.
This behavior has influence on the singularity structure of the
gluon propagator in the light-cone gauge $A^+=0$, notably,
$
 D_{\mu\nu}^{\rm LC} (q)
=
   \frac{-i}{q^2 - \lambda^2 + i0}
   \Big( g_{\mu\nu}
        -\frac{q_\mu n^-_\nu + q_\nu n^-_\mu}{[q^+]}
  \Big) ,
$
via the boundary conditions to go around its singularities.
To estimate this influence, one has to calculate the one-loop
virtual corrections in the $A^+=0$ gauge in conjunction with various
boundary conditions (which absorb large-scale effects) and carry out
the renormalization of the contour-dependent quark operators defined
in Eq.\ (\ref{eq:eikonal-quark}).
Two of us undertook this calculation, announced in
\cite{CS07,CS08,CS09}, with a summary of the approach being given in
\cite{SC09Trento}.
The contributing diagrams are shown here in Fig.\ \ref{fig:fig1},
while the corresponding algebraic expressions are given in Table
\ref{tab:virt-corr} using the following symbolic
abbreviations (the couplings $g$ and $g^{\prime}$ below are labeled
differently only in order to keep track of their origin; ultimately,
they will be set equal): \\
(i) $\mathbf{Q}$: struck quark
$
  \psi_i(\xi)
=
  e^{- ig [ \int\! d\eta \bar \psi \hat {\cal A} \psi ]}
$
$\psi_i^{\rm free} (\xi)$ --- Heisenberg operator, \\
(ii) longitudinal gauge link: $[n^-]$, \\
(iii) transverse gauge link: $[\mathbf{l}_{\perp}]$, \\
(iv) $g$ refers to the QCD Lagrangian --- see item (i), \\
(v) coupling $g^{\prime}$ refers to the exponent of the gauge links, i.e.,
                  $g^{\prime}\int_{0}^{\infty} d\tau \ldots$ \ , \\
(vi) product $g^{\prime}g^{\prime}$ corresponds to path-ordered
                  line integrals in the exponent of the gauge links, i.e.,
                  $g^{\prime}g^{\prime}\int_{0}^{\infty} d\tau \int_{0}^{\tau} d\sigma \ldots \ .$

\begin{table}[b]
\begin{tabular}{l|ccc}
                        & struck quark  & longitudinal gauge link  & transverse gauge link   \\
               \hline
struck quark            & $\mathbf{Q}\mathbf{Q}$ $\Longleftrightarrow$ (a)
                        & $\mathbf{Q}[n^-]$ ~~ $\Longleftrightarrow$ (b)
                        & $\mathbf{Q}[\mathbf{l}_{\perp}]$ ~~\, $\Longleftrightarrow$ (d)\\
longitudinal gauge link &
                        & $[n^-][n^-]$ $\Longleftrightarrow$ (c)
                        & $[n^-][\mathbf{l}_{\perp}]$ ~~~ $\Longleftrightarrow$ (f)=0               \\
transverse gauge link   &                                      &
                        & $[\mathbf{l}_{\perp}][\mathbf{l}_{\perp}]$ ~$\Longleftrightarrow$ (e)    \\
               \hline
\end{tabular}
\caption{Structure of the one-loop gluon virtual corrections to
$f_{q/q}(x, \mathbf{k}_{\perp})$ shown in Fig.\ 1.}
\label{tab:virt-corr}
\end{table}

Without going into too much detail, the results of this study show
that the overlapping ultraviolet (UV) and rapidity divergences cannot
be solely controlled by the dimensional (or any other) regularization.
The ensuing divergence is of the type $(1/\epsilon)\ln (\eta/p^{+})$,
which becomes infinite when $\eta \to 0$, and has, therefore, to be
cured by an appropriate renormalization procedure.
At this point it is important to mention that the terms on the
diagonal in Table \ref{tab:virt-corr} represent selfenergy
contributions, while all other terms are of the crosstalk type.
In the gauge $A^+=0$ only the terms $\mathbf{Q}\mathbf{Q}$ and
$\mathbf{Q}[\mathbf{l}_{\perp}]$ are non-vanishing.
Moreover, the pole-prescription dependence in diagram (a) is canceled
by its counterpart in (d) --- see Fig.\ \ref{fig:fig1}.
Taking into account the mirror contributions to (a) and (d) (not shown
in Fig.\ \ref{fig:fig1}), one finds the following total contribution
from virtual gluon corrections \cite{CS08,CS07}:
\begin{equation}
  \Sigma_{\rm UV}^{\rm (a+d)}(\alpha_s, \epsilon)
=
    2\frac{\alpha_s}{\pi}C_{\rm F} \left[ \frac{1}{\epsilon}
    \left( \frac{3}{4}
  + \ln \frac{\eta}{p^+} \right) - \gamma_E + \ln 4\pi \right] \ .
\label{eq:total-virt}
\end{equation}
From this expression one obtains for $f_{q/q}(x, \mathbf{k}_{\perp})$
the anomalous dimension $\left(\gamma
=
  \frac{\mu}{2}\frac{1}{Z}\frac{\partial\alpha_s}{\partial\mu}
  \frac{\partial Z}{\partial\alpha_s}
\right)$
\begin{equation}
  \gamma_{\rm one-loop}^{\rm LC}
=
  \frac{\alpha_s}{\pi} C_{\rm F}
  \left(
        \frac{3}{4} + \ln \frac{\eta}{p^{+}}
  \right)
=
  \gamma_{\rm smooth} -\delta\gamma \ ,
\label{eq:anom-dim}
\end{equation}
where $\eta$ is the rapidity parameter with $[\eta]=[\rm mass]$ and
$\delta\gamma$ represents the deviation from the anomalous
dimension of the gauge-invariant quark propagator in a covariant
gauge (see \cite{Ste83} and earlier references cited therein).
As argued in \cite{CS07,CS08,SC09Trento}, such an anomalous dimension
can be associated with a cusp in the gauge contour at infinity and
originates from the renormalization of the gluon interactions with this
local contour obstruction.
Therefore, one can claim that $\delta\gamma$ can be identified with the
universal cusp anomalous dimension \cite{KR87} at the one-loop order.
But the choice of the gauge $A^+=0$ should not affect the
renormalization properties of the TMD PDF.
Thus, the definition of $f_{q/q}(x, \mathbf{k}_{\perp})$ given by
Eq.\ (\ref{eq:tmd-pdf}) has to be modified by a soft factor
(counter term) \cite{CH00}
\begin{equation}
  R
\equiv
 \Phi (p^+, n^- | 0) \Phi^\dagger (p^+, n^- | \xi) \ ,
\label{eq:soft-factor}
\end{equation}
where $\Phi$ and $\Phi^\dagger$ are appropriate eikonal factors to be
evaluated along a jackknifed contour off the light cone (the explicit
expressions and a graphic illustration can be found in
\cite{CS07,CS08,SC09Trento}).
We have shown there by explicit calculation that in the $A^+=0$ gauge
with $q^-$-independent pole prescriptions (advanced, retarded,
principal value), the anomalous dimension associated with this quantity
exactly cancels $\delta\gamma$, rendering the modified definition of
the TMD PDF free from gauge artifacts.
On the other hand, adopting instead a $q^-$-dependent pole prescription
(Mandelstam \cite{Man82}, Leibbrandt \cite{Lei83}), no
anomalous-dimension anomaly appears and the soft factor reduces
benignly to unity \cite{CS09}.

\section{Inclusion of Pauli Spin Interactions}
\label{sec:Pauli}

The conventional way to restore the gauge invariance of hadronic matrix
elements is to use gauge links as those defined in
Eqs.\ (\ref{eq:lightlike-link}) and (\ref{eq:transverse-link}).
However, this is only the minimal way to achieve this goal; it ignores
the direct spin interactions because the gauge potential $A_{\mu}^{a}$
is spin-blind.
To accommodate the direct interaction of spinning particles with the
gauge field, one has to take into account the so-called Pauli term
$\sim F^{\mu\nu}S_{\mu\nu}$, where
$S_{\mu\nu}=\frac{1}{4}[\gamma_{\mu}, \gamma_{\nu}]$
is the spin operator.
Following this generalized conception of gauge invariance, we promote
the definition of the TMD PDF to \cite{CKS10}
\begin{eqnarray}
  f_{i/h}^{\Gamma}(x, \mathbf{k}_{\perp})
=
   \frac{1}{2} {\rm Tr} \int\! dk^-
   \int \frac{d^4 \xi}{(2\pi)^4}
   e^{- i k \cdot \xi}
   \langle  h |\bar \psi_i (\xi)
   [[
     \xi^{-}, {\bm \xi}_{\perp};\infty^{-}, {\bm \xi}_{\perp}
   ]]^\dagger
   [[
     \infty^-, {\bm \xi}_{\perp};\infty^-, {\bm \infty}_{\perp}
   ]]^\dagger \nonumber \\
   \times \Gamma
   {[[}
     \infty^-, {\bm \infty}_{\perp};\infty^{-}, \mathbf{0}_{\perp}
   {]]}
   [[\infty^-, \mathbf{0}_{\perp};
   0^-, \mathbf{0}_{\perp}]]
   \psi_i (0) | h
   \rangle
   \cdot R \  ,
\label{eq:TMD-PDF-Pauli}
\end{eqnarray}
where $\Gamma$ denotes one or more $\gamma$ matrices in correspondence
with the particular distribution in question, and the state
$|h\rangle$ stands for the appropriate target.
In the unpolarized case we have $|h\rangle=|h(P)\rangle$, with $P$
being the momentum of the initial hadron, whereas for a (transversely)
polarized target the state is
$|h\rangle=|h(P), S_{\perp}\rangle$.
The enhanced lightlike and transverse gauge links (denoted by double
square brackets) contain the Pauli term and are given, respectively,
by the following expressions:
\begin{eqnarray}
  [[\infty^-, \mathbf{0}_{\perp}; 0^-, \mathbf{0}_\perp]]
=
  \mathcal{P}
  \exp
      \left[
            - i g \!\!\! \int_{0}^{\infty} \! d\sigma \  u_{\mu}
                 A_{a}^{\mu}(u \sigma)t^a
            - i g \!\!\! \int_{0}^{\infty} \! d\sigma \
                 S_{\mu\nu} F_{a}^{\mu\nu}(u \sigma)t^a
      \right] \ ,
\end{eqnarray}
\begin{eqnarray}
  [[\infty^-,  {\bm \infty}_{\perp}; \infty^-, \mathbf{0}_\perp]]
=
  \mathcal{P}
  \exp
      \left[
            - i g \!\!\! \int_{0}^{\infty} \! d\tau
            \mathbf{l}_{\perp} \! \cdot \!
            \mathbf{A}_{\perp}^{a}(\mathbf{l}\tau)t^a
            - i g \!\!\! \int_{0}^{\infty} \!\!\! d\tau
            S_{\mu\nu}F_{a}^{\mu\nu}(\mathbf{l}\tau)t^a
      \right] \ .
\end{eqnarray}

\subsection{Gauge links with Pauli terms up to $\mathcal{O}(g^2)$}
\label{subsec:virtual-Pauli}

\begin{figure}[ht]
\centering
\includegraphics[width=0.4\textwidth,angle=90]{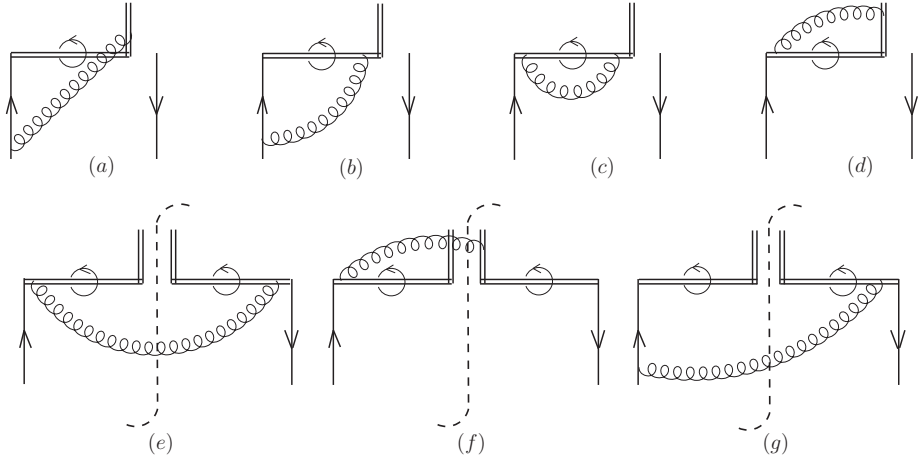}
\caption{One-loop gluon virtual corrections to $f_{q/q}$ in the
$A^+=0$ gauge.
Graphs (a), (b), (c), and (d) describe virtual gluon corrections;
graphs (e), (f), and (g) represent real-gluon exchanges
across the cut (vertical dashed line).
The double lines decorated with a ring represent enhanced gauge links
containing the Pauli term.}
\label{fig:pauli-virt-dia}
\end{figure}

Adopting this reasoning, we have to calculate in the $A^+=0$ gauge
the expression
\begin{equation}
  [[
    \infty^-, {\bm \infty}_\perp;\infty^-, \mathbf{0}_\perp
  ]]
\cdot
  [[\infty^-, \mathbf{0}_\perp;0^-, \mathbf{0}_\perp
  ]]
=
   1 - i g \left(
                 \mathcal{U}_{1} + \mathcal{U}_{2} + \mathcal{U}_{3}
         \right)
   - g^2 \left(
               \mathcal{U}_{4} + \mathcal{U}_{5}
                + \ldots \mathcal{U}_{10}
          \right)
\label{eq:gauge-links-product}
\end{equation}
with
$
 F_{a}^{\mu\nu} (\infty^-, 0^+, {\bm \xi}_\perp)
=
 0, ~~~
 \psi_i(\xi)
=
  e^{[- ig \int\! d\eta \ \bar \psi \hat {\cal A} \psi  ]}
  \psi_i^{\rm free} (\xi)
$
and an analogous expansion for the transverse gauge links
(see Table \ref{tab:link-terms}), whereas the contributing diagrams are
displayed in Fig.\ \ref{fig:pauli-virt-dia}.
\begingroup
\begin{table}[t]
\begin{tabular}{c l c l} 
Symbols & ~~~~~~~~~~~~~~~ Expressions & Figure 2 & ~~~ Value ~~~
\\ \hline
$\mathcal{U}_{1}$   &  $ \int_0^\infty \! d\tau \ \mathbf{l}
           \cdot \mathbf{{\cal A}}(\mathbf{l} \tau)$
       &  (a)
       &  $\neq 0$                                                        \\
$\mathcal{U}_{2}$   &  $ \int_0^\infty \! d\tau \ S \cdot {\cal F}(u \tau)$
       &  (b)
       &  $\neq 0$                                                        \\
$\mathcal{U}_{3}$   &  $ \int_0^\infty \! d\tau \ S \cdot {\cal F}(\mathbf{l} \tau)$
       &  ---
       &  $0$                                                             \\
$\mathcal{U}_{4}$   &  $\int_0^\infty \! d\tau \int_0^{\tau} d \sigma
        \ (\mathbf{l} \cdot \mathbf{{\cal A}}(\mathbf{l} \tau))
        \ (\mathbf{l} \cdot \mathbf{{\cal A}}(\mathbf{l} \sigma))
                       $
       & ---
       &  $0$                                                             \\
$\mathcal{U}_{5}$   &  $\int_0^\infty \! d\tau \int_0^{\tau} d \sigma
        \ (\mathbf{l} \cdot \mathbf{{\cal A}}(\mathbf{l} \tau))
        \ (S\cdot {\cal F}(\mathbf{l} \sigma))
          $
       &  ---
       &  $0$                                                             \\
$\mathcal{U}_{6}$   &  $\int_0^\infty \! d\tau \int_0^{\tau} d \sigma
        \ (S\cdot {\cal F}(\mathbf{l} \tau))\ (\mathbf{l}
          \cdot \mathbf{{\cal A}}(\mathbf{l} \sigma))
          $
       &  ---
       &  $0$                                                             \\
$\mathcal{U}_{7}$  &  $ \int_0^\infty \! d\tau \int_0^{\tau} d \sigma
        \ (S\cdot {\cal F}(u \tau))\ (S \cdot {\cal F}(u \sigma))
          $
       &  (c)
       &  $0$                                                             \\
$\mathcal{U}_{8}$ &  $\int_0^\infty \! d\tau \int_0^{\tau} d \sigma
       \  (S\cdot {\cal F}(\mathbf{l} \tau))\ (S \cdot {\cal F}(\mathbf{l} \sigma))
          $
       &  ---
       &  $0$                                                             \\
$\mathcal{U}_{9}$  &  $\int_0^\infty \! d\tau \int_0^{\infty} d \sigma
       \  (\mathbf{l} \cdot \mathbf{{\cal A}}(\mathbf{l} \tau))\
          (S\cdot {\cal F} (u \sigma) )$
       &  (d)
       &  $\neq 0$                                                        \\
$\mathcal{U}_{10}$  &  $\int_0^\infty \! d\tau \int_0^{\infty} d \sigma
       \  (S\cdot {\cal F}(\mathbf{l} \tau))\ (S \cdot {\cal F}(u \sigma))
          $
       &  ---
       & $0$                                                              \\
\hline
\end{tabular}
\caption{Individual virtual-gluon contributions appearing in
the evaluation of Eq.\ (\protect\ref{eq:gauge-links-product}) up to
$\mathcal{O}(g^2)$.}
\label{tab:link-terms}
\end{table}
\endgroup
Let us quote here some important features of the presented theoretical
framework referring for details to our recent work in Ref.\
\cite{CKS10}:
(i) The Pauli term is not reparameterization invariant --- unlike the
usual Dirac term.
Therefore, we have to use the dimensionful vectors
$
 n^+_\mu \to u^*_\mu
=
 p^- n^+_\mu \ , \quad
 n^-_\mu
\rightarrow
 u_\mu
=
 p^+ n^-_\mu \ , \quad
 \mathbf{l}_{\perp}
\rightarrow
 p^+ \mathbf{l}_{\perp}
$.
(ii) The Pauli spin-interaction terms do not completely vanish along
$n^-$ in the $A^+=0$ gauge, whereas terms containing
${\cal F} (\mathbf{l} \tau)$ (or ${\cal F} (\mathbf{l} \sigma)$)
cancel out in the product of the gauge links and
$
 F_{a}^{\mu\nu} (\infty^-, 0^+, {\bm \xi}_\perp)
=
  0
$.
(iii) To the $g^2$-order level, the Pauli term reads
\begin{equation}
  S \cdot {\cal F}
\equiv
  S_{\mu\nu} {\cal F}^{\mu \nu}
=
        2 S_{+-} {\cal F}^{+-}
     +  2 S_{+i} {\cal F}^{+i}
     +  2 S_{-i} {\cal F}^{-i}
     +  S_{ij} {\cal F}^{ij}
\end{equation}
and has the following non-zero components:
\begin{eqnarray}
  {\cal F}^{+-}
& = &
  \partial^+ {\cal A}^- \ , \ {\cal F}^{+i} = \partial^+ {\cal A}^i \ ,
\\
  \ {\cal F}^{-i}
& = &
  \partial^- {\cal A}^i - \partial^i {\cal A}^- \ , \
  {\cal F}^{ij}
=
  \partial^i {\cal A}^j - \partial^j {\cal A}^i \ .
\end{eqnarray}
(iv) The diagrams (a)--(d) in Fig\ \ref{fig:pauli-virt-dia}
represent virtual gluon corrections and contain UV and rapidity
divergences that give rise to the anomalous dimension of the TMD PDF.
In contrast, the diagrams (e)--(g), which describe real-gluon
exchanges across the cut (vertical dashed line), contribute only
finite terms.

From Fig.\ \ref{fig:pauli-virt-dia}, we see that the gauge-link
correlator contains contributions of two different types related to
selfenergy- and crosstalk-type diagrams.
To discuss the structure of the correlator in a compact way, it is
useful to use the following symbolic abbreviations: \\
$\displaystyle\mathbb{Q}$:  Gauge self-field in the Heisenberg quark operator
$
 \psi_i(\xi)
=
 e^{- ig[ \int\! d\eta \ \bar \psi \hat {\cal A} \psi ]} \
 \psi_i^{\rm free} (\xi)
$ \\
\vspace{0.2cm}
$\displaystyle
\mathbf{l} \cdot \mathbf{{\cal A}}(\mathbf{l} \tau)
\equiv
\mathbb{A^\perp}$: Standard transverse gauge potential \\
\vspace{0.2cm}
$\displaystyle
S \cdot {\cal F}(\mathbf{l} \tau)
\equiv
\mathbb{F}$: Tensor (Pauli) term

Then we obtain at $\mathcal{O}(g^2)$ the following results
(consult Fig.\ \ref{fig:pauli-virt-dia} in conjunction with Table
\ref{tab:link-terms}):

\textbf{Selfenergy-type contributions}
\begin{itemize}
\item
$\displaystyle\mathbb{A}^\perp \displaystyle\mathbb{A}^\perp$:
$
 \langle \mathcal{U}_{4} \rangle
=
 0 $ ~~~~~ not shown
\item
$\mathbb{F^-} \mathbb{F^-}$~:
$
 \langle \mathcal{U}_{7} \rangle
=
 0 $ ~~~~~ diagram (c) in Fig.\ \ref{fig:pauli-virt-dia}
\item
$\mathbb{F^\perp} \mathbb{F^\perp}$~:
$
 \langle \mathcal{U}_{8} \rangle
=
 0 $ ~~~~~ not shown
\end{itemize}

\textbf{Crosstalk-type contributions}
\begin{itemize}
\item
$
 \displaystyle\mathbb{Q} \mathbb{A^\perp}
$:
$
 \langle \mathcal{U}_{1} \rangle^{\rm UV}
=
  - \alpha_s \, C_{\rm F}\, \frac{1}{\varepsilon}\, i \, C_\infty
$
with
$C_\infty = \{ 0 (\rm adv); -1 (ret); -\frac{1}{2} (PV) \}$
--- diagram (a).
This term cancels the pole-prescription-dependent term in the UV-divergent
part of the fermion selfenergy
$\displaystyle\mathbb{Q} \displaystyle\mathbb{Q}$.
\item
$
 \displaystyle\mathbb{Q} \mathbb{F^-}
$:
$
 \langle \mathcal{U}_{2} \rangle
$
with
$
 (\displaystyle\mathbb{Q} \mathbb{F^-})^-
=
 \langle \mathcal{U}_{2}^- \rangle
 $
and
$
 (\displaystyle\mathbb{Q} \mathbb{F^-})^\perp
=
 \langle \mathcal{U}_{2}^\perp \rangle
$ --- diagram (b).
Accordingly, for the leading twist-two TMD PDF, we find for the
semi-inclusive DIS (SIDIS)
\begin{eqnarray}
  \Gamma_{\rm tw-2} \langle \mathcal{U}_{2}^- \rangle
  + \langle \mathcal{U}_{2}^- \rangle^\dagger \Gamma_{\rm tw-2}
& = &
  \frac{i}{2} C_{\rm F} \Gamma_{\rm tw-2} \ , \\
\Gamma_{\rm tw-2} \langle \mathcal{U}_{2}^\perp \rangle
  +  \langle \mathcal{U}_{2}^\perp \rangle^\dagger \Gamma_{\rm tw-2}
& = &
     -\frac{i}{4} C_{\rm F} \ \Gamma_{\rm tw-2} \, .
\label{eq:U-2}
\end{eqnarray}
These two results combine to produce a constant phase
(unrelated to that found in \cite{BJY03})
\begin{equation}
  \delta_{\rm tw-2} = \alpha_s C_{\rm F} \pi
\label{eq:phase}
\end{equation}
which is also valid for the twist-three TMD PDF, i.e.,
$
 \delta_{\rm tw-3} = \alpha_s C_{\rm F} \pi
$,
but \emph{flips sign} for the Drell-Yan (DY) process
because it depends on the direction of the longitudinal gauge link.
Hence, our analysis \cite{CKS10} predicts the important relation
\begin{equation}
  \delta_{\rm SIDIS} = - \delta_{\rm DY} \, .
\label{eq:DY-phase}
\end{equation}
\item
$
 \displaystyle\mathbb{Q} \mathbb{F^\perp}
$:
$
 \langle \mathcal{U}_{3} \rangle
=
 0
$ ~~~~~ not shown
\item
$
 \mathbb{A^\perp} \mathbb{F^\perp}
$:
$
 \langle \mathcal{U}_{6} \rangle
=
 - \langle \mathcal{U}_{5} \rangle
$, hence mutually canceling
\item
$
 \mathbb{A^\perp} \mathbb{F^-}
$:
$
 \langle \mathcal{U}_{9} \rangle
=
 \langle \mathcal{U}_{9} \rangle^\dagger
$
--- diagram (d) in Fig.\ \ref{fig:pauli-virt-dia} ---
(``gluon mass'' $\lambda^2$ drops out at the end):
\begin{equation}
  \langle \mathcal{U}_{9} \rangle
=
  - \frac{1}{8\pi} \
  C_{\rm F} [\gamma^+, \gamma^-] \Gamma (\epsilon)
  \left( 4\pi \frac{\mu^2}{\lambda^2} \right)^\epsilon \ .
\label{eq:U-9}
\end{equation}
This nontrivial Dirac structure entails
\begin{eqnarray}
& &  \Gamma_{\rm unpol.}
 =
  \gamma^+ ~~~~~~~\! :
  \Gamma_{\rm unpol.} [\gamma^+, \gamma^-]
  =
    - [\gamma^+, \gamma^-] \Gamma_{\rm unpol.} \ , \\
& &  \Gamma_{\rm helic.}
\;  = \gamma^+ \gamma^5
    ~~~           :
    \, \Gamma_{\rm helic.} [\gamma^+, \gamma^-]
~\!  =
    - [\gamma^+, \gamma^-]\Gamma_{\rm helic.}  \ , \\
& &  \Gamma_{\rm trans.}
\;  =  i \sigma^{i+} \gamma^5 :
    \, \Gamma_{\rm trans.} [\gamma^+, \gamma^-]
~\! =
    - [\gamma^+, \gamma^-] \Gamma_{\rm trans} \, ,
\end{eqnarray}
where obvious acronyms have been used.
Taking into account the mirror diagrams
(not shown in Fig.\ \ref{fig:pauli-virt-dia}),
the twist-two terms mutually cancel by virtue of the relation
$$
  [\gamma^+, \gamma^-] \Gamma_{\rm tw-2}
=
  - \Gamma_{\rm tw-2} [\gamma^+, \gamma^-]
=
  2 \Gamma_{\rm tw-2} \ ,
$$
which permits a probabilistic interpretation of the twist-two TMD PDF
as a density on account of
$
 \mathbb{A^\perp} \mathbb{F^-}\rightarrow 0
$.
On the other hand, the twist-three TMD PDF gets a non-vanishing
contribution to its anomalous dimension as one sees from
$$
 \Gamma_{\rm tw-3} \langle \mathcal{U}_{9} \rangle
+
  \langle \mathcal{U}_{9} \rangle^{\dagger} \Gamma_{\rm tw-3}
=
 -\frac{C_{\rm F}}{4\pi} [\gamma^+, \gamma^-] \Gamma (\epsilon)
  \left(\! 4\pi \frac{\mu^2}{\lambda^2}\! \right)^\epsilon \ .
$$
\item
$
 \mathbb{F^\perp} \mathbb{F^-}
$:
$
 \langle \mathcal{U}_{10} \rangle
=
 0
$
without assuming any particular form of the gauge field at light cone
$\infty$.
\end{itemize}

\subsection{Real-Gluon Contributions at $\mathcal{O}(g^2)$}
\label{subsec:real-Pauli}
Besides the virtual gluon corrections, there are also real gluon
exchanges that contribute finite contributions to the TMD PDF.
The main difference from the previously considered case is that now
the discontinuity goes across the gluon propagator that has to be
replaced by the cut one.
Moreover, the Dirac structures, marked above by the symbol $\Gamma$,
are sandwiched between Dirac matrices stemming from Pauli terms
standing on different sides of the cut.
The real-gluon contributions are specified in Table 3.

\begingroup
\begin{table}[t]
\begin{tabular}{c l c } 
Symbols & ~~~~~~~~~~~~~~~ Expressions & Figure \protect\ref{fig:pauli-virt-dia}
\\ \hline
$\mathcal{U}_{11}$   &  $ \int_0^\infty \! d\tau \int_0^{\infty} d \sigma
        \ (S \cdot {\cal F}(u \tau))\ \Gamma \
          (S \cdot {\cal F}(u \sigma + \xi^-; {\bm \xi}_\perp))
          $
       &  (e) \\
$\mathcal{U}_{12}$  &  $\int_0^\infty \! d\tau \int_0^{\infty} d \sigma
       \  (\mathbf{l} \cdot {\cal A}(\mathbf{l} \tau))\ \Gamma \
          (S \cdot {\cal F}(u \sigma + \xi^-; {\bm \xi}_\perp))
          $
       &  (f) \\
$\mathcal{U}_{13}$ &  $ \int_0^{\infty} d \sigma
              \Gamma \ (S\cdot {\cal F}
              (u \sigma + \xi^-; {\bm \xi}_\perp) )
          $
       &  (g)                      \\
\hline
\end{tabular}
\caption{Individual real-gluon contributions to $\mathcal{O}(g^2)$
corresponding to the diagrams $(e), (f), (g)$ in
Fig.\ \protect\ref{fig:pauli-virt-dia}.}
\label{tab:link-terms_real}
\end{table}
\endgroup
Using the same symbolic notation as in the previous subsection, we
briefly remark that
\begin{itemize}
\item
$
 \mathbb{F^-} \mathbb{F^-}
$:
$
 \langle \mathcal{U}_{11} \rangle
\rightarrow 0$ (at least power-suppressed $\sim p^-$)
\item
$
 \mathbb{A^\perp}  \mathbb{F^-}
$:
$
 \langle \mathcal{U}_{12} \rangle + \langle \mathcal{U}_{12} \rangle^\dagger
\sim
  \Gamma [\gamma^+, \gamma^-] + [\gamma^+, \gamma^-] \Gamma
=
 0
$
\item
$
 \displaystyle\mathbb{Q} \mathbb{F^-}
$:
$
 \langle \mathcal{U}_{13}^- \rangle
  + \langle \mathcal{U}_{13}^- \rangle^\dag
$
and
$
 \langle \mathcal{U}_{13}^\perp \rangle
  + \langle \mathcal{U}_{13}^\perp \rangle^\dag
$
mutually cancel up to a power-suppressed term.
\end{itemize}

\section{Highlights and Conclusions}
\label{sec:concl}
\vspace{-0.1cm}

We argued that the dimensional regularization of overlapping UV and
rapidity divergences in TMD PDFs is not sufficient to render the TMD
PDF finite --- one needs renormalization \cite{CS07,CS08}.
To remedy this deficiency, a soft factor \cite{CH00} along a jackknifed
contour off the light cone was introduced into the definition of the
TMD PDF \cite{CS07} whose anomalous dimension cancels in leading loop
order the cusp anomalous dimension entailed by this overlapping
divergence (with a full-fledged discussion being given in \cite{CS08}).
The modified TMD PDF reproduces the standard integrated PDF and is
controlled by an evolution equation with the same anomalous dimension
as one finds in covariant gauges with no dependence on the adopted pole
prescription for the gluon propagator --- this would be impossible
without the soft renormalization factor (see \cite{CS08} and for a more
dedicated discussion \cite{CS09Yer}).
In particular, using the $A^+=0$ gauge in conjunction with the
Mandelstam-Leibbrandt pole prescription \cite{Man82,Lei83}, no
anomalous-dimension defect appears and thus the soft factor becomes
trivial.
An important finding of this approach is that the anomalous dimension
of the unpolarized TMD PDF for SIDIS and the DY process is the same,
i.e.,
$\gamma_{f_{q/q}}^{\rm SIDIS}=\gamma_{f_{q/q}}^{\rm DY}$,
albeit the sign of the $\epsilon$ term in the gluon propagator
$
 \frac{1}{q^+ + i \epsilon}
$
is different for these two processes --- irrespective
of the boundary condition applied.
Quite recently, Collins discussed alternative ways to redefine the
TMD PDFs in such a way as to avoid rapidity divergences \cite{Col02}.

We also presented a new scheme for gauge-invariant TMD PDFs which
includes the direct interaction of spinning particles with the
gauge field by means of the Pauli term in the longitudinal and
transverse gauge links.
In some sense, the Pauli spin interaction is the abstract analogue of
a Stern-Gerlach apparatus --- sort of --- and gives rise through the
transverse gauge link to a constant phase
$\delta=\alpha_s C_{\rm F} \pi$, which is the same for twist-two
and twist-three TMD PDFs, but flips sign when the direction of the
gauge link is reversed --- thus breaking universality.
As a result, one finds
$\delta_{\rm DY} = -\delta_{\rm SIDIS}$.
To facilitate calculations, we developed in Ref.\ \cite{CKS10} Feynman
rules for enhanced gauge links --- longitudinal and transverse ---
which supplement those derived before for the standard gauge links by
Collins and Soper \cite{CS81}.
Because the Pauli term contributes to the anomalous dimension of the
twist-three TMD PDF, the evolution of such quantities is more delicate
and may require the modification of the renormalization factor to
preserve its density interpretation.

Bottom line: Our results --- most significant amongst them the
appearance of a non-universal phase --- may stimulate both theoretical
and experimental activities.
On the other hand, T-even and T-odd TMD PDFs may become ``measurable''
on the lattice, so that it seems possible that non-trivial Wilson
lines, as those we discussed in this presentation, may be revealed in
the future.

\acknowledgments
We are grateful to Anatoly Efremov for useful discussions.
We acknowledge financial support from the Heisenberg--Landau Program,
Grant 2010, the DAAD, and the INFN.
N.G.S. is thankful to the DAAD for a travel grant and to the Organizers
for the hospitality and financial support.


\begin{thebibliography}{99}
\bibitem{CS81}
  J.C.~Collins and D.E.~Soper,
  \emph{Back-to-back jets in QCD},
  \emph{Nucl.\ Phys.} {\bf B193} (1981) 381;
  \emph{Nucl.\ Phys.} {\bf B213} (983) 545 (E).

\bibitem{CS08}
  I.O.~Cherednikov and N.G.~Stefanis,
  \emph{Wilson lines and transverse-momentum dependent parton distribution
  functions: A renormalization-group analysis},
  \emph{Nucl.\ Phys.} {\bf B802} (2008) 146
  [{\tt arXiv:0802.2821 [hep-ph]}].

\bibitem{Ste83}
  N.G.~Stefanis,
  \emph{Gauge Invariant Quark Two Point Green's Function Through
  Connector Insertion To $\mathcal{O}(\alpha_s)$},
  \emph{Nuovo Cim.} {\bf A83} (1984) 205.

\bibitem{Man68YM}
  S.~Mandelstam,
  \emph{Feynman rules for electromagnetic and Yang-Mills fields
        from the gauge independent field theoretic formalism},
  \emph{Phys.\ Rev.} {\bf 175} (1968) 1580.

\bibitem{BJY03}
  A.V.~Belitsky, X.~Ji and F.~Yuan,
  \emph{Final state interactions and gauge invariant parton distributions},
  \emph{Nucl.\ Phys.} {\bf B656} (2003) 165
   [{\tt arXiv:hep-ph/0208038}].

\bibitem{BMP03}
  D.~Boer, P.J.~Mulders and F.~Pijlman,
  \emph{Universality of T-odd effects in single spin and azimuthal asymmetries},
  \emph{Nucl.\ Phys.} {\bf B667} (2003) 201
  [{\tt arXiv:hep-ph/0303034}].

\bibitem{CS07}
  I.O.~Cherednikov and N.G.~Stefanis,
  \emph{Renormalization, Wilson lines, and transverse-momentum dependent parton
  distribution functions},
  \emph{Phys.\ Rev.} {\bf D77} (2008) 094001
  [{\tt arXiv:hep-ph/0710.1955}].

\bibitem{CS09}
  I.O.~Cherednikov and N.G.~Stefanis,
  \emph{Renormalization-group properties of transverse-momentum dependent parton
  distribution functions in the light-cone gauge with the Mandelstam-Leibbrandt
  prescription},
  \emph{Phys.\ Rev.} {\bf D80} (2009) 054008
  [{\tt arXiv:hep-ph/0904.2727}].

\bibitem{SC09Trento}
  N.G.~Stefanis and I.O.~Cherednikov,
  \emph{Renormalization-group anatomy of transverse-momentum dependent parton
  distribution functions in QCD},
  \emph{Mod.\ Phys.\ Lett.} {\bf A24} (2009) 2913
  [{\tt arXiv:hep-ph/0910.3108}].

\bibitem{KR87}
  G.P.~Korchemsky and A.V.~Radyushkin,
  \emph{Renormalization of the Wilson Loops Beyond the Leading Order},
  \emph{Nucl.\ Phys.} {\bf B283} (1987) 342.

\bibitem{CH00}
  J.C.~Collins and F.~Hautmann,
  \emph{Infrared divergences and non-lightlike eikonal lines in Sudakov processes},
  \emph{Phys.\ Lett.} {\bf B472} (2000) 129
  [{\tt hep-ph/9908467}].

\bibitem{Man82}
  S.~Mandelstam,
  \emph{Light Cone Superspace And The Ultraviolet Finiteness Of The N=4 Model},
  \emph{Nucl.\ Phys.} {\bf B213} (1983) 149.

\bibitem{Lei83}
  G.~Leibbrandt,
  \emph{The Light Cone Gauge In Yang-Mills Theory},
  \emph{Phys.\ Rev.} {\bf D29} (1984) 1699.

\bibitem{CKS10}
  I.O.~Cherednikov, A.I.~Karanikas and N.G.~Stefanis,
  \emph{Wilson lines in transverse-momentum dependent parton distribution functions
  with spin degrees of freedom},
  \emph{Nucl.\ Phys.} {\bf B840} (2010) 379
   [{\tt arXiv:1004.3697 [hep-ph]}].

\bibitem{CS09Yer}
  I.O.~Cherednikov and N.G.~Stefanis,
  \emph{Understanding the evolution of transverse-momentum dependent parton
  densities},
  Talk given at Workshop on Transverse Partonic Structure of Hadrons
  (TPSH 2009), Yerevan, Armenia, 21-26 Jun 2009
   [{\tt arXiv:0911.1031 [hep-ph]}].

\bibitem{Col02}
  J.~Collins,
  \emph{Rapidity divergences and valid definitions of parton densities},
  in proceedings of
  \emph{LIGHT CONE 2008 Relativistic Nuclear and Particle Physics},
  July 7-11, 2008, Mulhouse, France,
  \pos{PoS(LC2008)028}
  [{\tt arXiv:0808.2665 [hep-ph]}].

\end{thebibliography}
\end{document}